----------------------------------------------------------------------------
\date{} 
\documentstyle{article} 
\begin{document} 
\begin{center} {\bf \Large $SU(3)_{flavor}$ analysis of
two-body weak decays of charmed baryons} \\  
\vskip 0.5 cm
\large   K. K. Sharma and R. C. Verma
\protect\footnotemark[1]\footnotetext[1]{Present address:  Department of
Physics, Punjabi University, Patiala-147 002, India}\\ \normalsize
Centre for Advanced Study in Physics, Department of Physics,\\
\normalsize Panjab University, Chandigarh -160014  India\\ 
\vskip 0.5cm
{ABSTRACT}  
\end{center}\large
\baselineskip 24pt 
\hskip 0.5truecm 
\par We study two-body weak decays
of charmed  baryons $\Lambda^{+}_{c},$ $\Xi^{+}_{c}$ and $\Xi^{0}_{c}$
into an octet or decuplet baryon and a pseudoscalar meson employing the
$SU(3)$ flavor symmetry. Using certain measured Cabibbo-favored modes,
we fix the reduced amplitudes and predict the branching ratios of
various decays of charmed baryons in the Cabibbo-enhanced, -suppressed
and -doubly suppressed modes.
\vskip 0.5cm
\par 
\noindent PACS numbers(s): 13.30.Eg, 11.30.Hv, 11.40.Ha, 14.20.Lq
\newpage 
\large  {\bf I. INTRODUCTION}  
\par \baselineskip 24pt 
\par 
As more new data [1-4] on charmed baryons become available in recent years,
the theoretical study of nonleptonic weak decays of charmed baryons has
acquired significance. Earlier, it was hoped that like meson decays the
spectator quark process would dominate for charm baryon decays also.
However, this scheme does not seem to be supported by experiment, as the
observed branching ratio for decays like  $\Lambda^{+}_{c} \rightarrow
\Sigma^{+} \pi^{0}/\Xi^{0} K^{+}$, forbidden in the spectator quark
model, are significantly large thereby indicating the need of
nonspectator contributions. Generally, these contributions are treated
through the current algebra approach and soft pion techniques [5].
Unfortunately, the calculations of both pole terms and factorizable
contributions have their own uncertainties associated with many
parameters and even by adjusting all the parameters, agreement with the
experimental observations is far from satisfactory [6].

An alternative to the above approach is to employ flavor symmetry
approach [7-9]. Though, this approach involves a number of unknown
reduced amplitudes, it has the advantage that it lumps all the dynamical
processes together. In contrast to the badly broken $SU(4)$ charm
scheme, $SU(3)$ flavor symmetry is expected to be more reliable for the
study of charm baryons. Recently, one of us (RCV) and Khanna [10] have
studied the Cabibbo-favored $(CF)$ decays of charmed baryons in the
$SU(3)$ flavor symmetry generated by u, d and s quarks. In this work, we
extend this approach to study Cabibbo-suppressed $(CS)$ and
-doubly-suppressed $(CDS)$  $B_c \rightarrow BP/DP$ decays (where $B_c$
represents the charmed baryon and $B/D$ the octet/decuplet baryon and
$P$ a pseudoscalar meson respectively). Using the data available on
$\Lambda^{+}_{c} \rightarrow \Lambda \pi^{+}/\Sigma^{+} \pi^{0}/\Xi^{0}
K^{+}$ decays, we determine the reduced amplitudes, which are then used
to predict branching ratios and asymmetry of various CF, CS and CDS
decays. Similarly,  we also study $B_c \rightarrow DP$ decays, where we
use $Br(\Lambda^+_c \rightarrow \Delta^{++}K^-/\Xi^{*0}K^+)$ to fix the
reduced amplitudes. \par {\bf II. THEORETICAL FRAMEWORK}  \par Structure
of the general weak current $\otimes$ current Hamiltonian $H_{w}$
including short distance $QCD$ effects for the charm changing
Cabibbo-favored decays $(\Delta C = \Delta S = -1)$ is $$H_w~=~\tilde
{G_F}[c_{1}(\bar u d)(\bar s c) + c_{2}(\bar s d)(\bar u c)], \eqno(1)
$$ where $\tilde {G_F}~=~\frac {G_F}{\sqrt2}V_{ud}V^*_{cs}.$  $ \bar
q_{1} q_{2} \equiv \bar q_{1} \gamma_{\mu} (1 - \gamma_{5} ) q_{2} $
represents color singlet $ V - A $ current and the $QCD$ coefficients at
the charm mass scale are $$c_{1}~~=~~1.26 \pm 0.04,~~~~~ c_{2}~~=~~-0.51
\pm 0.05. \eqno(2)$$ The effective weak Hamiltonian $(1)$ transforms as
an admixture of the $6^*$ and $15$ representations of the
$SU(3)$-flavor, which can be expressed as $$H^{6^{*}}_{W}~~=~~\sqrt2
g_{8_{S}} \{\bar B^{a}_{m}P^{m}_{b}B^{n}H^{b}_{[n,a]} + \bar
B^{m}_{b}P^{a}_{m}B^{n}H^{b}_{[n,a]}\}$$ $$\\ ~~~~~~~~+ \sqrt2 g_{8_{A}}
\{\bar B^{a}_{m}P^{m}_{b}B^{n}H^{b}_{[n,a]} - \bar
B^{m}_{b}P^{a}_{m}B^{n}H^{b}_{[n,a]}\}$$ $$ \\ ~~~~~+ \frac {\sqrt2}{2}
g_{10^{*}} \{\bar B^{a}_{b}P^{c}_{d}B^{b}H^{d}_{[a,c]} + \bar
B^{a}_{b}P^{c}_{d}B^{d}H^{b}_{[a,c]}$$ $$ \\ ~~~~~- \frac {1}{3}\bar
B^{a}_{b}P^{c}_{a}B^{n}H^{b}_{[n,c]} + \frac {1}{3}\bar
B^{a}_{c}P^{c}_{d}B^{n}H^{d}_{[n,a]}\}, \eqno(3)$$ $$H^{15}_{W}~~=~~
\frac {\sqrt2}{2} h_{27} \{\bar B^{a}_{b}P^{c}_{d}B^{b}H^{d}_{(a,c)} +
\bar B^{a}_{b}P^{c}_{d}B^{d}H^{b}_{(a,c)}$$ $$ \\ ~~~~~- \frac
{1}{5}\bar B^{a}_{b}P^{c}_{a}B^{n}H^{b}_{(n,c)} - \frac {1}{5}\bar
B^{a}_{c}P^{c}_{d}B^{n}H^{d}_{(n,a)}\}$$ $$\\ ~~~~~+ \frac {\sqrt2}{2}
h_{10} \{\bar B^{a}_{b}P^{c}_{d}B^{b}H^{d}_{(a,c)} - \bar
B^{a}_{b}P^{c}_{d}B^{d}H^{b}_{(a,c)}$$ $$ \\ ~~~~~+ \frac {1}{3}\bar
B^{a}_{b}P^{c}_{a}B^{n}H^{b}_{(n,c)} - \frac {1}{3}\bar
B^{a}_{c}P^{c}_{d}B^{n}H^{d}_{(n,a)}\}$$ $$\\  ~~~~~+ \sqrt2 h_{8_{S}}
\{\bar B^{a}_{m}P^{m}_{b}B^{n}H^{b}_{(n,a)} + \bar
B^{m}_{b}P^{a}_{m}B^{n}H^{b}_{(n,a)}\}$$ $$\\ ~~~~~~~~+ \sqrt2 h_{8_{A}}
\{\bar B^{a}_{m}P^{m}_{b}B^{n}H^{b}_{(n,a)} - \bar
B^{m}_{b}P^{a}_{m}B^{n}H^{b}_{(n,a)}\}, \eqno(4)$$ where the $QCD$
coefficients $c_{1}$ and $c_{2}$ get absorbed in the reduced amplitudes
$g's$ and $h's$. Here, $B^{a}~~\equiv~~(-\Xi^{0}_{c},\hskip 0.1truecm
\Xi^{+}_{c},\hskip 0.1truecm \Lambda^{+}_{c}),$ and $B^a_b$ denote the
antitriplet of charmed baryons and octet baryons respectively.
$P^{a}_{b}$ denotes $3 \times 3$ matrix of the uncharmed pseudoscalar
meson nonet $$ P^{a}_{b} \hskip 0.5cm = \hskip 0.5cm \left(\matrix {
P^{1}_{1} & \pi^{+} & K^{+} \cr \pi^{-} & P^{2}_{2} & K^{0} \cr K^{-} &
\bar K^{0} & P^{3}_{3} \cr}\right), \eqno(5)$$ with $$ P^{1}_{1} \hskip
0.5cm = \hskip 0.5cm \frac {1}{ \sqrt2 }\{ \pi^{0} ~~+~~ \eta ~sin
\theta ~~+~~ \eta'~cos \theta \},$$ $$ P^{2}_{2} \hskip 0.5cm = \hskip
0.5cm \frac {1}{ \sqrt2 }\{-~~ \pi^{0} ~~+~~ \eta ~sin \theta ~~+~~
\eta'~cos \theta \},$$   $$ P^{3}_{3} \hskip 0.5cm = \hskip 0.5cm \{-~~
\eta ~cos \theta ~~+~~ \eta'~sin \theta \}, \eqno(6)$$ where $\theta$
governs the $\eta-\eta'$ mixing and is related to the physical mixing as
$$ \theta ~~~=~~~ \theta_{ideal}~~ - ~~ \phi_{phy}. \eqno(7)$$ \par The
amplitude for the decay process $B_{c} \rightarrow BP$ is defined by
$$<B_{f}P|H_{W}|B_{i}>~~~=~~~ i\bar u_{B_{f}}\{A~-~\gamma
_{5}B\}u_{B_{i}} \phi_{p},\eqno(8)$$ where $A$ and $B$ are respectively
$s-$wave and $p-$wave amplitudes and $u_{B}$ are the Dirac spinors. This
gives the decay rate $$ \Gamma (B_{i} \rightarrow B_{f}~+~P)~~~=~~~
C_{1} \{ |A|^{2}~+~C_{2}|B|^{2}\}, \eqno(9)$$ and asymmetry parameter $$
\alpha ~~~=~~~ \frac {2~Re(A\bar B^{*})}{(|A|^{2}~~+~~|\bar B|^{2})},
\eqno(10)$$ with $\bar B~~=~~\sqrt C_{2} B.$ The kinematical factors
$C_{1},$ $C_{2}$ are given by $$C_{1}~~~=~~~\frac {|p_{c}|}{8\pi} \frac
{(m_{i}~+~m_{f})^{2}~~-~~m^{2}_{p}}{m^{2}_{i}}, \eqno(11)$$
$$C_{2}~~~=~~~ \frac
{(m_{i}~-~m_{f})^{2}~~-~~m^{2}_{p}}{(m_{i}~+~m_{f})^{2}~~+~~m^{2}_{p}},
\eqno(12)$$ where $p_{c}$ is the center of mass three-momentum in the
rest frame of the parent particle. $m_i$, $m_f$ are masses of initial
and final state baryons respectively and $m_p$ is the mass of the meson
emitted. \par  \noindent {\bf III. NUMERICAL CALCULATIONS AND RESULTS}
\par {\bf A. Cabibbo-favored mode} \par To illustrate the procedure, we
discuss the main steps involved in the determination of the reduced
amplitudes. Taking $H^{2}_{13}$ component of the weak Hamiltonian in (3)
and (4), the decay amplitudes of various Cabibbo-favored decays of
antitriplet charmed baryons are obtained [7, 8, 10]. There are seven
reduced amplitudes in each of the $PV$ and $PC$ modes.  Assuming $6^*$
dominance of the weak Hamiltonian, we reduce the number of unknown
parameters from seven to three in each of these modes. \par A recent
$CLEO$ measurement [3] has reported the following set of $PV$ and $PC$
amplitudes (in the units of $~G_{F} V_{ud} V^*_{cs}~ \times ~10^{-2}~
GeV^{2})$ $$A(\Lambda^+_c \rightarrow \Lambda
\pi^+)~~=~~-3.0^{+0.8}_{-1.2}~~or~~-4.3^{+0.8}_{-0.9},$$ $$B(\Lambda^+_c
\rightarrow \Lambda
\pi^+)~~=~~+12.7^{+2.7}_{-2.5}~~or~~+8.9^{+3.4}_{-2.4},$$
$$A(\Lambda^+_c \rightarrow \Sigma^{+}
\pi^0)~~=~~+1.3^{+0.9}_{-1.1}~~or~~+5.4^{+0.9}_{-0.7},$$ $$B(\Lambda^+_c
\rightarrow \Sigma^{+}
\pi^0)~~=~~-17.3^{+2.3}_{-2.9}~~or~~-4.1^{+3.4}_{-3.0}. \eqno(13)$$ It
has been shown [10] that the present data on $Br(\Lambda^{+}_{c}
\rightarrow p \bar K^{0})$ prefers the following set: $$A(\Lambda^+_c
\rightarrow \Lambda \pi^+)~~=~~-3.0^{+0.8}_{-1.2},~~~B(\Lambda^+_c
\rightarrow \Lambda \pi^+)~~=~~+12.7^{+2.7}_{-2.5},$$ $$A(\Lambda^+_c
\rightarrow \Sigma^{+} \pi^0)~~=~~+5.4^{+0.9}_{-0.7}, ~~~B(\Lambda^+_c
\rightarrow \Sigma^{+} \pi^0)~~=~~-4.1^{+3.4}_{-3.0}. \eqno(14)$$

Further, experimental branching ratio $Br(\Lambda^{+}_{c} \rightarrow
\Xi^{0} K^{+})~ =~ (0.34 \pm 0.09)\%$ [9] yields, $$|A(\Lambda^{+}_{c}
\rightarrow \Xi^{0} K^{+})|^{2}~~+~~C_2|B(\Lambda^{+}_{c} \rightarrow
\Xi^{0} K^{+})|^{2}~~~=~~~14.42\pm 3.82. \eqno(15)$$ Various dynamical
mechanisms considered for the charm baryon decays indicate that the PV
mode of this decay is highly suppressed. This decay in $PV$ mode can
neither occur through the spectator quark scheme nor from the equal time
commutator $(ETC)$ term of the current algebra. Even through the $(\frac
{1}{2})^-$ baryon pole, it acquires a negligibly small contribution
[11]. Therefore, taking $\alpha (\Lambda^{+}_{c} \rightarrow \Xi^{0}
K^{+})~ \approx ~ 0$, we fix $$|B(\Lambda^{+}_{c} \rightarrow \Xi^{0}
K^{+})|~= ~ \pm (16.52\pm 2.19). \eqno(16) $$Using the decay amplitudes of
$\Lambda^+_c \rightarrow \Lambda \pi^+/\Sigma^+ \pi^0/\Xi^0K^+$, we
express the reduced amplitudes as follows: $$ g_{8_S}~=~\frac {1}{2}\{
\frac {1}{ \sqrt2}<\Sigma^+ \pi^0|\Lambda^+_c>-\sqrt {\frac {3}{2}}
<\Lambda \pi^+|\Lambda^+_c>-<\Xi^0K^+|\Lambda^+_c>\},\eqno(17)$$ $$
g_{8_A}~=~\frac {1}{6}\{ \frac {5}{ \sqrt2}<\Sigma^+
\pi^0|\Lambda^+_c>-\sqrt { \frac {3}{2}} <\Lambda
\pi^+|\Lambda^+_c>+<\Xi^0K^+|\Lambda^+_c>\},\eqno(18)$$ $$ g_{10}~=~\{
\frac {1}{\sqrt2}<\Sigma^+ \pi^0|\Lambda^+_c>+\sqrt {\frac {3}{2}}
<\Lambda \pi^+|\Lambda^+_c>-<\Xi^0K^+|\Lambda^+_c>\},\eqno(19)$$ which
can be used to determine the other decays. For instance, $\Lambda^+_c
\rightarrow p\bar K^0$ decay amplitude is expressed as: $$<p\bar
K^0|\Lambda^+_c> = \frac {1}{\sqrt2}~\{ \sqrt3 <\Lambda
\pi^+|\Lambda^+_c>-<\Sigma^+ \pi^0|\Lambda^+_c>\},$$
$$~~~~~~~~~~~~=~-7.49\pm 1.35 ~for~ PV~ mode,$$
$$~~~~~~~~~~~~=~+18.45\pm 3.91~for ~PC~ mode, \eqno(20) $$ where the
error is calculated using the average of errors given in (14). Thus we
calculate $$Br(\Lambda^+_c \rightarrow p\bar K^0)~=~(2.67\pm 0.74)\%,
\eqno(21) $$ and $$\alpha ~=~ -0.99\pm 0.39, \eqno(22)$$ which agrees
with the observed experimental branching ratio of $(2.1\pm 0.4)\%$ [1].
Following this procedure, we determine branching ratio and asymmetry of
remaining Cabibbo-enhanced decays taking negative and positive signs of
$B(\Lambda^+_c \rightarrow \Xi^0 K^+)$. For $\Lambda_c \rightarrow
\Sigma \pi$ decays, isospin symmetry yields: $$Br(\Lambda^+_c
\rightarrow \Sigma^0 \pi^+)~=~Br(\Lambda^+_c \rightarrow \Sigma^+
\pi^0),\eqno(23)$$ $$ (0.87\pm0.20)\%~~~=~~~(0.87\pm 0.22)\% \hskip
0.5cm (Expt.)$$ which holds well, and $$\alpha (\Lambda^+_c \rightarrow
\Sigma^0 \pi^+)~=~\alpha (\Lambda^+_c \rightarrow \Sigma^+
\pi^0)~=~(-0.45\pm 0.31\pm 0.06). \eqno(24) $$ For $\Lambda^+_c
\rightarrow \Sigma^+ \eta,$ we obtain $$Br(\Lambda^{+}_{c} \rightarrow
\Sigma^{+} \eta)~=~(0.50\pm 0.17)\%,~ at ~ \phi_{phy} = -10^0,$$
$$~~~~~~~~~~~~~~~~~~~~~=~(0.55\pm 0.19)\%,~ at ~ \phi_{phy} =
-19^0,\eqno(25)$$ with the negative sign of $B(\Lambda^{+}_{c}
\rightarrow \Xi^{0} K^{+})$, and $$Br(\Lambda^{+}_{c} \rightarrow
\Sigma^{+} \eta)~=~(0.97\pm 0.23)\%,~ at ~ \phi_{phy} = -10^0,$$
$$~~~~~~~~~~~~~~~~~~~=~~~(1.23\pm 0.28)\%,~ at ~ \phi_{phy} =
-19^0,\eqno(26)$$ with the positive sign. A recent $CLEO$ measurement
[2] $$\frac {Br(\Lambda^+_c \rightarrow \Sigma^+ \eta)}{Br(\Lambda^+_c
\rightarrow p K^- \pi^+)}~=~0.11 \pm 0.03 \pm 0.02, \eqno(27)$$ combined
with $Br(\Lambda^+_c \rightarrow p K^- \pi^+)~=~4.4 \pm 0.6\%$ [1]
yields $$ Br(\Lambda^{+}_{c} \rightarrow \Sigma^{+} \eta)~~=~~0.48 \pm
0.17,\eqno(28)$$ which seems to prefer the negative sign of
$B(\Lambda^{+}_{c} \rightarrow \Xi^{0} K^{+}).$ Recently, branching
ratio of $\Xi^{+}_{c} \rightarrow \Xi^{0} \pi^{+}$ has also been
measured in a CLEO-II experiment [4] to be $(1.2\pm 0.5 \pm 0.3)\% .$
For this mode we obtain values $(4.14\pm 1.27)\%$ and $(0.07\pm 0.02)\%$
for negative and positive signs of $B(\Lambda^{+}_{c} \rightarrow
\Xi^{0} K^{+})$ respectively. Thus experiment seems to prefer the
positive sign. Therefore, in Tables I(a) and I(b), we give branching
ratios of CF decays for both the signs, for the sake of comparison. The
decays $\Xi^0_c \rightarrow \Lambda \bar K^0/\Sigma^0\bar
K^0/\Xi^-\pi^+$ remain unaffected by the sign of  $B(\Lambda^{+}_{c}
\rightarrow \Xi^{0} K^{+}).$ \par \noindent {\bf B. Cabibbo-suppressed
mode} \par Effective weak Hamiltonian for these decays  $(\Delta C =
-1$, $\Delta S = 0)$ is given by $$ H_w~=~ \tilde {G'_F}[c_{1} \{(\bar u
d)(\bar d c)-(\bar u s)(\bar s c)\}+c_{2} \{(\bar d d)(\bar u c)-(\bar s
s)(\bar u c)\}], \eqno(29) $$ where $\tilde {G'_F}~=~-\frac
{G_F}{\sqrt2}V_{ud}V^*_{cd}$ and other quantities have the usual
meanings. Choosing $(H^{2}_{12}-H^{3}_{13})$ components of the weak
Hamiltonian in (3) and (4), decay amplitudes for various
Cabibbo-suppressed decays are obtained [7, 8]. As the same reduced
amplitudes appear here, the CS decay amplitudes can be expressed in
terms of those of the CF modes. In the following, we obtain some of
these relations using $6^*$ dominance of $H_w$

$$- tan \theta_{c} <\Xi^{0} K^{+}|\Lambda^{+}_{c}>~~~=~~~-~<pK^{-}
|\Xi^{0}_{c}>~~~=~~~<\Sigma^{+} \pi^{-} |\Xi^{0}_{c}>, \eqno(30)$$
$$\sqrt2 ~tan \theta_{c} ~<\Sigma^{+}
\pi^{0}|\Lambda^{+}_{c}>~~~=~~~-~<n \bar K^{0}
|\Xi^{0}_{c}>~~~=~~~<\Xi^{0} K^{0} |\Xi^{0}_{c}>, \eqno(31)$$
$$- tan \theta_{c}\{ \sqrt {\frac {3}{2}} <\Lambda
\pi^+|\Lambda^{+}_{c}>-\frac {1}{\sqrt2}<\Sigma^+
\pi^0|\Lambda^+_c>\}~~~=~~~~<\Sigma^{-} \pi^{+} |\Xi^{0}_{c}>$$
$$~~~~~~~~~~~~~~~~~~~~~~~=~~-~<\Xi^{-} K^{+} |\Xi^{0}_{c}>, \eqno(32)$$
$$- tan \theta_{c} \{\frac {1}{\sqrt2} <\Sigma^{+}
\pi^{0}|\Lambda^{+}_{c}>+ \sqrt {\frac {3}{2}} <\Lambda
\pi^+|\Lambda^{+}_{c}>\}~~~=~~~-\sqrt2~<p \pi^{0}|\Lambda^{+}_{c}>$$
$$~~~~~~~~~~~=~~~-~<n \pi^{+}|\Lambda^{+}_{c}>
~~~~~=~~<\Xi^0K^+|\Xi^+_c>,\eqno(33)$$
$$- tan \theta_{c} \{-\sqrt2 <\Sigma^{+}
\pi^{0}|\Lambda^{+}_{c}>+<\Xi^{0} K^{+} |\Lambda^{+}_{c}>\}
~~~=~~~-\sqrt2~<\Sigma^{0} K^{+}|\Lambda^{+}_{c}>$$
$$~~~~~~~~~~~~~~~~~=~~~-~< \Sigma^{+}K^{0}|\Lambda^{+}_{c}>
~~~=~~~<p\bar K^0 |\Xi^{+}_{c}>, \eqno(34)$$
$$- tan \theta_{c} \{~<\Xi^{0} K^{+}|\Lambda^{+}_{c}>-\sqrt {\frac
{2}{3}}~<\Lambda \pi^{+} |\Lambda^{+}_{c}>\}~=~\sqrt {\frac
{2}{3}}~<\Lambda K^{+} |\Lambda^{+}_{c}>,\eqno(35)$$
$$- tan \theta_{c} \{\sqrt {\frac {3}{2}} <\Lambda
\pi^+|\Lambda^{+}_{c}>-\frac {1}{\sqrt2}<\Sigma^+
\pi^0|\Lambda^+_c>+<\Xi^{0} K^{+} |\Lambda^{+}_{c}>\}$$
$$~~~~~~~~~~~~~~~~~~~~~~~~=~~~2~<\Sigma^{0} \pi^{0} |\Xi^{0}_{c}>,
\eqno(36)$$
$$- tan \theta_{c} \{-~<\Xi^{0} K^{+}|\Lambda^{+}_{c}>~+~\sqrt {\frac
{3}{2}} <\Lambda \pi^{+} |\Lambda^{+}_{c}>~- ~\frac
{1}{\sqrt2}~<\Sigma^+\pi^0|\Lambda^+_c>\}$$
$$~~~~~~~~~~~=~\sqrt2~<\Sigma^{0}
\pi^{+}|\Xi^{+}_{c}>~~~=~~~-\sqrt2~<\Sigma^{+} \pi^{0}|\Xi^{+}_{c}>,
\eqno(37)$$
$$- tan \theta_{c} \{3~<\Xi^{0} K^{+}|\Lambda^{+}_{c}>-\sqrt {\frac
{3}{2}}~<\Lambda \pi^{+} |\Lambda^{+}_{c}>-\frac {3}{\sqrt2}~<\Sigma^{+}
\pi^{0} |\Lambda^{+}_{c}>\}$$ $$~~~~~~~~~~~~~=~~~\sqrt6~<\Lambda
\pi^{+}|\Xi^{+}_{c}>~~~=~~~-\sqrt{12}~<\Lambda \pi^{0}|\Xi^{0}_{c}>,
\eqno(38)$$ We give the decay asymmetries and branching ratios for the
$CS$ decays, in Tables II(a) and II(b) for both the signs of
$B(\Lambda^{+}_{c} \rightarrow \Xi^{0} K^{+}).$ In the present analysis,
we find that the decays $\Xi^+_c \rightarrow p\bar K^0/\Lambda \pi^+$
and $\Xi^0_c \rightarrow \Sigma^- \pi^+/\Xi^- K^+$ are dominant for both
choices. Among the $\Lambda^+_c$ decays, $\Lambda^+_c \rightarrow
\Lambda K^+/p\eta$ and $ \Lambda^+_c \rightarrow  \Sigma^+K^0$ are
dominant for negative and positive signs of $B(\Lambda^{+}_{c}
\rightarrow \Xi^{0} K^{+})$ respectively.

\par \noindent {\bf C. Cabibbo-doubly-suppressed mode} \par For the
Cabibbo-doubly-suppressed decays  $(\Delta C = -\Delta S =-1) $ the
effective weak Hamiltonian is $$H_w~=~ \tilde {G''_F}[c_{1} (\bar u
s)(\bar d c) + c_{2}(\bar d s)(\bar u c)], \eqno(39) $$ where $\tilde
{G''_F}~=~-\frac {G_F}{\sqrt2}V_{us}V^*_{cd}$. Here also the $CDS$
decays can be expressed in term of the $CF$ modes. Using $6^*$ dominance
of the weak Hamiltonian, we obtain the following decay amplitude
relations:$$- tan^{2} \theta_{c}  ~< \Xi^{0}
K^{+}|\Lambda^{+}_{c}>~~~=~~~\sqrt2~<p \pi^{0}|\Xi^{+}_{c}>~~~=~~~<n
\pi^{+} |\Xi^{+}_{c}>$$ $$~~~~~~~~~~~~\\ ~~~=~~~-~\sqrt2~<n
\pi^{0}|\Xi^{0}_{c}>~~~=~~~<p \pi^{-} |\Xi^{0}_{c}>, \eqno(40)$$ $$-
tan^{2} \theta_{c}  ~\{\sqrt {\frac {3}{2}}~<\Lambda \pi^{+}
|\Lambda^{+}_{c}>-\frac {1}{\sqrt2}~<\Sigma^{+} \pi^{0}
|\Lambda^{+}_{c}>\}~~~=~~~\sqrt2~<\Sigma^{0} K^{+}|\Xi^{+}_{c}>$$
$$~~~~~=~~~<\Sigma^{+} K^{0}|\Xi^{+}_{c}> ~~~=~~~<\Sigma^{-} K^{+}
|\Xi^{0}_{c}>~~~=~~~-~\sqrt2~<\Sigma^{0} K^{0}|\Xi^{0}_{c}>, \eqno(41)$$
$$- tan^{2} \theta_{c} \{ -\frac {3}{\sqrt2}~<\Sigma^{+}
\pi^{0}|\Lambda^{+}_{c}>-\sqrt {\frac {3}{2}}~<\Lambda \pi^{+}
|\Lambda^{+}_{c}>\}~~~=~~~\sqrt6~<\Lambda K^{+}|\Xi^{+}_{c}>$$
$$~~~~~~~~~~~~~~~~~~~~~=~~~\sqrt{6}~<\Lambda K^{0}|\Xi^{0}_{c}>,
\eqno(42)$$
$$- tan^{2} \theta_{c} \{ -\sqrt {\frac {3}{2}}~<\Lambda \pi^{+}
|\Lambda^{+}_{c}>-\frac {1}{\sqrt2}~<\Sigma^{+} \pi^{0}
|\Lambda^{+}_{c}>~ +~ < \Xi^{0} K^{+}|\Lambda^{+}_{c}>~\}$$
$$~~~~~~~~~~~~~~~~\\
~~~=~~~-~<pK^{0}|\Lambda^{+}_{c}>~~~=~~~~<nK^{+}|\Lambda^{+}_{c}>.
\eqno(43)$$

Calculated asymmetries and branching ratios of the $CDS$ decays are
given in the Tables III(a) and III(b). Among $\Xi_{c}$ decays, $\Xi^+_c
\rightarrow \Sigma^+ K^0/\Sigma^0 K^+/n\pi^+/p\pi^0$ and $\Xi^0_c
\rightarrow \Sigma^- K^+$ are found to be dominant modes for positive as
well as negative choice of $B(\Lambda^{+}_{c} \rightarrow \Xi^{0}
K^{+}).$ However, branching ratios of $\Lambda^+_c$ decays show drastic
difference between the two choices, even their decay asymmetries also
acquire different signs.

\newpage \par  \noindent ${\bf IV.~ B_c(\frac {1}{2})^+ \rightarrow
D(\frac {3}{2})^+~+~P(0^-)~ Decays }$  \par \noindent The matrix element
for the baryon  $(\frac {1}{2})^+ \rightarrow (\frac {3}{2})^+~+~ 0^-$
decay process is $$ M~=~<D,P|H_w|B_c>~=~iP_\mu \bar {w}^\mu_D(C -
\gamma_5 D)u_{B_c}\phi_P, \eqno(44) $$ where $P_\mu$ is the four
momentum of the meson and $w^\mu$ is the Rarita-Schwinger spinor for a
spin $3/2^+$ particle. C and D denote the $p-wave$ and $d-wave$
amplitudes respectively. The decay rate and asymmetry parameter are
computed from $$ \Gamma (B_{i} \rightarrow B_{f}~+~P)~~~=~~~ \frac {
|p_c|^3 m_i(m_f + E_f)}{6\pi m^2_f}\{ |C|^{2}~+~|\bar D|^{2}\},
\eqno(45)$$ $$ \alpha ~~~=~~~ \frac {2~Re(C\bar
D^{*})}{(|C|^{2}~~+~~|\bar D|^{2})}, \eqno(46)$$ where $\bar D$ is
defined as $$ \bar D~=~\rho D, ~~ \rho ~= ~ \{ \frac {E_f - m_f}{E_f +
m_f} \}^{1/2}. \eqno(47)$$ $E_f$ is the energy of the final state baryon
in the rest frame of $B_c$ and other quantities have the usual meaning.
The weak Hamiltonian for decuplet baryon emitting decays is given by $$
H^{6^*}_W ~~ = ~~ \sqrt2 j_8\{ \epsilon_{mdb} \bar {D}^{mnc}P^d_n B^a
H^b_{[a,c]}\}, \eqno(48)$$ $$H^{15}_W~=~\sqrt2 k_8 \{ \epsilon_{mpb}
\bar {D}^{mna}P^p_n B^c H^b_{(a,c)}\}$$ $$~~~~~~~~~~~+ \sqrt2 k_{10} \{
\epsilon_{mnd} \bar {D}^{mac}P^n_b B^d H^b_{(a,c)} - \epsilon_{mnb} \bar
{D}^{mac}P^n_d B^d H^b_{(a,c)}$$ $$~~~~~~~~~~~+\frac {2}{3}
\epsilon_{mnb} \bar {D}^{mdc}P^n_d B^a H^b_{(a,c)}\}$$ $$~~~~~~~~~~~+
\sqrt2 k_{27} \{ \epsilon_{mnd} \bar {D}^{mac}P^n_b B^d H^b_{(a,c)} +
\epsilon_{mnb} \bar {D}^{mac}P^n_d B^d H^b_{(a,c)}$$ $$~~~~~~~~~~~-\frac
{2}{5} \epsilon_{mnb} \bar {D}^{mdc}P^n_d B^a H^b_{(a,c)}\}, \eqno(49)$$
where $\epsilon_{abc}$ is the Levi-Civita symbol and $ D_{abc}$
represents the totally symmetric decuplet baryons. \vskip 0.5cm \par
Decay amplitudes for CF, CS, and CDS modes are obtained by taking
$H^2_{13},$ $(H^2_{12}-H^3_{13}),$ and $H^3_{12}$  components of the
weak Hamiltonian [8, 10]. Here, we have four unknown reduced amplitudes
in each of the $PV$ and $PC$ modes. Dynamically, in contrast to
$B(\frac {1}{2})^+ \rightarrow B(\frac {1}{2})^+~+~ P(0)^-$ decays, the
description of $B(\frac {1}{2})^+ \rightarrow D(\frac {3}{2})^+~+~
P(0)^-$ is considerably simpler. It has been shown [12] that the prime
feature of these decays is that, they are factorization forbidden and
arise only through W-exchange diagrams. Also Kohra [13], while
performing a quark-diquark analysis, has observed that most of the quark
diagrams, allowed for  $(\frac {1}{2})^+ \rightarrow (\frac {1}{2})^+~+~
0^-$ decays are forbidden for $(\frac {1}{2})^+ \rightarrow (\frac
{3}{2})^+~+~ 0^-$ decays due to the symmetry property of the decuplet
baryons. There exist only two independent diagrams which correspond to
$$A ~=~ d_1\bar {D}^{1ab}B_{[2,a]}M^3_b +  d_2\bar
{D}^{3ab}B_{[2,a]}M^1_b. \eqno (50) $$ This amounts to the following
constraints: $$k_8~=~\frac {1}{3}k_{10}, ~~k_{27}~=~0, \eqno(51)$$ for
the 15-part of the weak Hamiltonian in  our model. Thus, the number of
unknown reduced amplitudes is reduced to two $(j_8$ and $k_8)$.
Generally, the W-exchange diagram contributions to the PV mode are small
and it is invariably suppressed due to the centrifugal barrier for $B
\rightarrow D + P $ decays. Therefore, we ignore them in the present
analysis. Experimental values [1] $$Br(\Lambda^{+}_c \rightarrow
\Delta^{++} K^-)~=~(0.7\pm 0.4)\%,\eqno(52)$$ $$Br(\Lambda^{+}_c
\rightarrow \Xi^{*0} K^+)~=~(0.23\pm 0.09)\%, \eqno(53)$$ then yields
(in $G_F V_{ud}V^*_{cs}~ \times ~10^{-2}~GeV^{2})$ $$k_{8}~=~-9.10\pm
4.15, ~~j_8~=~-77.14\pm 12.45. \eqno(54)$$ Using these, we calculate the
branching ratios, which are listed in column (ii) of Tables IV, V and
VI, for Cabibbo-enhanced, -suppressed and -doubly suppressed modes
respectively. In the Cabibbo-enhanced mode, $\Lambda^+_c \rightarrow
\Sigma^{*+} \pi^0 / \Sigma^{*0} \pi^+$ and $\Xi^0_c \rightarrow \Xi^{*-}
\pi^+/ \Omega^- K^+$ dominate, whereas $\Xi^+_c$ decays remain forbidden
in the present model like other theoretical models. In the $CS$ sector,
we find that the decays $\Lambda^+_c \rightarrow
\Delta^+\pi^0/\Delta^0\pi^+,~ \Xi^+_c \rightarrow
\Delta^{++}K^-/\Sigma^{*+}\pi^0/\Sigma^{*0}\pi^+$ and $\Xi^0_c
\rightarrow \Sigma^{*-}\pi^+$ are dominant. In the $CDS$ mode,
$\Lambda^+_c \rightarrow \Delta^+K^0/\Delta^0K^+$ decays are forbidden,
and $\Xi^+_c \rightarrow \Delta^{++}\pi^-/\Delta^+\pi^0/\Delta^0\pi^+, ~
\Xi^0_c \rightarrow \Delta^-\pi^+/\Delta^0\pi^0$ decays are dominant. We
hope that the observation of these decays would decipher the strength of
various weak decay mechanisms, particularly of the 15-part of weak
Hamiltonian. \vskip 0.5cm \par  \noindent {\bf V. ~ Summary and
discussion}  \par \noindent The two-body weak decays of charmed baryons
$\Lambda^{+}_{c}$, $\Xi^{+}_{c}$ and $\Xi^{0}_{c}$ into an octet or
decouplet baryon and a pseudoscalar meson are analysed in the framework
of $SU(3)$ flavor symmetry, for Cabibbo-enhanced, -suppressed and
doubly-suppressed modes. We fix the unknown reduced amplitudes from
certain measured Cabibbo-enhanced modes and then predict the branching
ratios and asymmetries of various decays. This work was motivated by the
observation that various dynamical models used for studying these decays
are far from explaining the data on $\Lambda_{c}$-decays. In the flavor
symmetry approach various processes responsible for the decays are
lumped together in the reduced amplitudes. However, the results obtained
here, may be affected by the $SU(3)$ symmetry breaking, as is evident
from the the charm meson decays [14] and the $\Lambda_c$, and
$\Xi^{0}_{c}$ lifetimes [1]. In the present framework , the inclusion of
the $SU(3)$ symmetry breaking effects would introduce a large number of
parameters which can not be determined with the available data.

\begin{center} \newpage  \large Table I(a) Branching ratios and
 asymmetries of CF $(B_c \rightarrow BP)$ decays for $B(\Lambda^+_c
 \rightarrow \Xi^0K^+) = -16.52 \pm 2.19$ \baselineskip 24pt
 \begin{tabular}{ | c|c|c| }  \hline Decay & Asymmetry & $ Br\%$ \\
 \hline & & \\ $ \Lambda^{+}_{c} \rightarrow p \bar K^{0}$ &$-0.99\pm
 0.39$  &$2.67\pm 0.74$ \\ $ \Lambda^{+}_{c} \rightarrow \Lambda
 \pi^{+}$ &$-0.94\pm 0.24^*$ &$0.79\pm0.18^*$ \\ $ \Lambda^{+}_{c}
 \rightarrow \Sigma^+ \pi^0$ &$-0.45\pm 0.32^*$ &$0.87\pm 0.22^*$ \\ $
 \Lambda^{+}_{c} \rightarrow \Sigma^+\eta $ &$0.92\pm 0.47^1 ~(0.76 \pm
 0.43^2)$ &$0.50\pm 0.17^1 ~(0.55 \pm 0.19^2)$ \\ $ \Lambda^{+}_{c}
 \rightarrow \Sigma^+\eta' $ &$-0.75\pm 0.38^1 ~(-0.89 \pm 0.46^2)$
 &$0.20\pm 0.08^1~(0.16 \pm 0.06^2)$  \\ $ \Lambda^{+}_{c} \rightarrow
 \Sigma^0 \pi^+$ &$-0.45\pm 0.32$ &$0.87\pm 0.20$  \\ $ \Lambda^{+}_{c}
 \rightarrow \Xi^0 K^+$ &$-0.00$ &$0.34\pm 0.09^*$ \\ & & \\ $
 \Xi^{+}_{c} \rightarrow \Xi^0 \pi^+$ &$0.03\pm 0.31$ &$4.14\pm 1.27$
 \\ $ \Xi^{+}_{c} \rightarrow \Sigma^+ \bar K^0$ &$0.03\pm 0.29$
 &$4.18\pm 1.28$\\& & \\ $ \Xi^{0}_{c} \rightarrow \Xi^0 \pi^0$
 &$0.72\pm 0.41$ &$0.52\pm 0.15$ \\ $ \Xi^{0}_{c} \rightarrow \Xi^0
 \eta$ &$-0.96\pm 0.38^1~(-0.95 \pm 0.32^2)$ &$0.29\pm 0.08^1~(0.37 \pm
 0.08^2)$ \\ $ \Xi^{0}_{c} \rightarrow \Xi^0 \eta'$ &$-0.63\pm
 0.40^1~(-0.60 \pm 0.48^2)$ &$0.12\pm 0.05^1~(0.08 \pm 0.04^2)$ \\ $
 \Xi^{0}_{c} \rightarrow \Xi^- \pi^+$ &$-0.96\pm 0.38$ &$1.30\pm 0.36$
 \\ $ \Xi^{0}_{c} \rightarrow \Sigma^+ K^-$ &$-0.00$ &$0.38\pm 0.10$ \\
 $ \Xi^{0}_{c} \rightarrow \Sigma^0 \bar K^0$ &$0.07\pm 0.67$ &$0.11\pm
 0.07$ \\ $ \Xi^{0}_{c} \rightarrow \Lambda \bar K^0$ &$-0.85\pm 0.36$
 &$0.68\pm 0.49$  \\ & & \\ \hline \end{tabular} \\ $^*$Input,  $^1$For
 $\phi_{phy}=-10^0$,  $^2$For $\phi_{phy}=-19^0$ \\

\newpage  Table I(b) Branching ratios and asymmetries of  CF $(B_c
\rightarrow BP)$ decays for $B(\Lambda^+_c \rightarrow \Xi^0K^+) =
+16.52 \pm 2.19$ \baselineskip 24pt \begin{tabular}{ | c|c|c| } \hline
Decay & Asymmetry & $ Br\%$ \\ \hline & & \\ $ \Lambda^{+}_{c}
\rightarrow p \bar K^{0}$ &$-0.99\pm 0.39$  &$2.67\pm 0.74$ \\ $
\Lambda^{+}_{c} \rightarrow \Lambda \pi^{+}$ &$-0.94\pm 0.24^*$
&$0.79\pm0.18^*$ \\ $ \Lambda^{+}_{c} \rightarrow \Sigma^+ \pi^0$
&$-0.45\pm 0.32^*$ &$0.87\pm 0.22^*$ \\ $ \Lambda^{+}_{c} \rightarrow
\Sigma^+\eta $ &$-0.96\pm 0.34^1~(-0.96 \pm 0.32^2)$ &$0.97\pm
0.23^1~(1.23 \pm 0.28^2)$ \\ $ \Lambda^{+}_{c} \rightarrow \Sigma^+\eta'
$ &$-0.91\pm 0.40^1~(-0.90 \pm 0.45^2)$ &$0.24\pm 0.08^1~(0.16 \pm
0.06^2)$  \\ $ \Lambda^{+}_{c} \rightarrow \Sigma^0 \pi^+$ &$-0.45\pm
0.32$ &$0.87\pm 0.20$  \\ $ \Lambda^{+}_{c} \rightarrow \Xi^0 K^+$
&$0.00$ &$0.34\pm 0.09^*$ \\ & & \\ $ \Xi^{+}_{c} \rightarrow \Xi^0
\pi^+$ &$-0.24\pm 0.23$ &$0.07\pm 0.02$  \\ $ \Xi^{+}_{c} \rightarrow
\Sigma^+ \bar K^0$ &$-0.23\pm 0.22$ &$0.07\pm 0.02$\\& & \\ $
\Xi^{0}_{c} \rightarrow \Xi^0 \pi^0$ &$-0.99\pm 0.37$ &$0.78\pm 0.20$ \\
$ \Xi^{0}_{c} \rightarrow \Xi^0 \eta$ &$0.14\pm 0.34^1~(-0.25\pm
0.29^2)$ &$0.19\pm 0.06^1~(0.25 \pm 0.07^2)$ \\ $ \Xi^{0}_{c}
\rightarrow \Xi^0 \eta'$ &$-0.99\pm 0.42^1~(-0.99 \pm 0.47^2)$ &$0.18\pm
0.06^1~(0.15\pm 0.05^2)$ \\ $ \Xi^{0}_{c} \rightarrow \Xi^- \pi^+$
&$-0.96\pm 0.38$ &$1.30\pm 0.36$ \\ $ \Xi^{0}_{c} \rightarrow \Sigma^+
K^-$ &$0.00$ &$0.38\pm 0.10$ \\ $ \Xi^{0}_{c} \rightarrow \Sigma^0 \bar
K^0$ &$0.07\pm 0.67$ &$0.11\pm 0.07$ \\ $ \Xi^{0}_{c} \rightarrow
\Lambda \bar K^0$ &$-0.85\pm 0.36$ &$0.68\pm 0.49$  \\ & & \\ \hline
\end{tabular} \\

$^{*}$Input,  $^1$For $\phi_{phy}=-10^0$,  $^2$For $\phi_{phy}=-19^0$

\newpage \large  Table II(a) Branching ratios and asymmetries of CS
$(B_c \rightarrow BP)$  decays  for $B(\Lambda^+_c \rightarrow \Xi^0K^+)
= -16.52 \pm 2.19$.\\ \baselineskip 24pt \begin{tabular}{ | c|c|c| }
\hline Decay & Asymmetry & $ Br\%$ \\ \hline & & \\ $ \Lambda^{+}_{c}
\rightarrow p \pi^{0}$ &0.05 &0.02 \\ $ \Lambda^{+}_{c} \rightarrow n
\pi^{+}$ &0.05 &0.04 \\ $ \Lambda^{+}_{c} \rightarrow \Lambda K^{+}$
&-0.54 &0.14 \\ $ \Lambda^{+}_{c} \rightarrow \Sigma^{+} K^{0}$ &0.68
&0.09 \\ $ \Lambda^{+}_{c} \rightarrow \Sigma^{0} K^{+}$ &0.68 &0.04 \\
$ \Lambda^{+}_{c} \rightarrow p\eta $ &$-0.74^1~(-0.69^2)$
&$0.21^1~(0.17^2)$ \\ $ \Lambda^{+}_{c} \rightarrow p\eta' $
&$-0.97^1~(-0.99^2)$ &$0.04^1~(0.06^2)$ \\ & & \\ $ \Xi^{+}_{c}
\rightarrow p\bar K^{0} $ &0.87 &0.19 \\ $ \Xi^{+}_{c} \rightarrow
\Lambda \pi^{+} $ &0.65 &0.23 \\ $ \Xi^{+}_{c} \rightarrow \Xi^{0} K^{+}
$ &0.08 &0.03 \\ $ \Xi^{+}_{c} \rightarrow \Sigma^{+} \pi^{0} $ &-0.89
&0.28 \\ $ \Xi^{+}_{c} \rightarrow \Sigma^{0} \pi^{+} $ &-0.90 &0.28 \\
$ \Xi^{+}_{c} \rightarrow \Sigma^{+} \eta $ &$-0.75^1~(-0.81^2)$
&$0.19^1~(0.21^2)$ \\ $ \Xi^{+}_{c} \rightarrow \Sigma^{+} \eta' $
&$-0.56^1~(-0.14^2)$ &$0.02^1~(0.02^2)$ \\ & & \\ $ \Xi^{0}_{c}
\rightarrow p K^{-} $ &-0.00 &0.03 \\ $ \Xi^{0}_{c} \rightarrow n \bar
K^{0} $ &-0.58 &0.04 \\ $ \Xi^{0}_{c} \rightarrow \Lambda \pi^{0} $
&0.65 &0.03 \\ $ \Xi^{0}_{c} \rightarrow \Sigma^{+} \pi^{-} $ &-0.00
&0.03 \\ $ \Xi^{0}_{c} \rightarrow \Sigma^{0} \pi^{0} $ &-0.18 &0.01 \\
$ \Xi^{0}_{c} \rightarrow \Sigma^{-} \pi^{+} $ &-0.99 &0.08 \\ $
\Xi^{0}_{c} \rightarrow \Xi^{-} K^{+} $ &-0.92 &0.06 \\ $ \Xi^{0}_{c}
\rightarrow \Xi^{0} K^{0} $ &-0.40 &0.04 \\ $ \Xi^{0}_{c} \rightarrow
\Lambda \eta $ &$0.26^1~(0.83^2)$ &$0.005^1~(0.003^2)$ \\ $ \Xi^{0}_{c}
\rightarrow \Lambda \eta' $ &$-0.82^1~(-0.77^2)$ &$0.02^1~(0.02^2)$ \\ $
\Xi^{0}_{c} \rightarrow \Sigma^{0} \eta $ &$-0.75^1~(-0.81^2)$
&$0.03^1~(0.03^2)$ \\ $ \Xi^{0}_{c} \rightarrow \Sigma^{0} \eta' $
&$-0.56^1~(-0.14^2)$ &$0.003^1~(0.002^2)$  \\ & & \\ \hline
\end{tabular} \\ $^1$For $\phi_{phy}=-10^0$,  $^2$For $\phi_{phy}=-19^0$
\\ \newpage  Table II(b) Branching ratios and asymmetries of CS $(B_c
\rightarrow BP)$ decays  for $B(\Lambda^+_c \rightarrow \Xi^0K^+) =
+16.52 \pm 2.19$\\  \baselineskip 24pt \begin{tabular}{ | c|c|c| }
\hline Decay & Asymmetry & $ Br\%$ \\ \hline & & \\ $ \Lambda^{+}_{c}
\rightarrow p \pi^{0}$ &0.05 &0.02 \\ $ \Lambda^{+}_{c} \rightarrow n
\pi^{+}$ &0.05 &0.04 \\ $ \Lambda^{+}_{c} \rightarrow \Lambda K^{+}$
&0.97 &0.02 \\ $ \Lambda^{+}_{c} \rightarrow \Sigma^{+} K^{0}$ &-0.98
&0.12 \\ $ \Lambda^{+}_{c} \rightarrow \Sigma^{0} K^{+}$ &-0.98 &0.06 \\
$ \Lambda^{+}_{c} \rightarrow p\eta $ &$-0.45^1~(-0.03^2)$
&$0.04^1~(0.02^2)$ \\ $ \Lambda^{+}_{c} \rightarrow p\eta' $
&$-0.99^1~(-0.99^2)$ &$0.05^1~(0.06^2)$ \\ & & \\ $ \Xi^{+}_{c}
\rightarrow p\bar K^{0} $ &-0.98 &0.36 \\ $ \Xi^{+}_{c} \rightarrow
\Lambda \pi^{+} $ &-0.79 &0.14 \\ $ \Xi^{+}_{c} \rightarrow \Xi^{0}
K^{+} $ &0.08 &0.03 \\ $ \Xi^{+}_{c} \rightarrow \Sigma^{+} \pi^{0} $
&-0.18 &0.08 \\ $ \Xi^{+}_{c} \rightarrow \Sigma^{0} \pi^{+} $ &-0.18
&0.08 \\ $ \Xi^{+}_{c} \rightarrow \Sigma^{+} \eta $
&$-0.98^1~(-0.98^2)$ &$0.08^1~(0.11^2)$ \\ $ \Xi^{+}_{c} \rightarrow
\Sigma^{+} \eta' $ &$-0.99^1~(-0.99^2)$ &$0.05^1~(0.03^2)$ \\  & & \\ $
\Xi^{0}_{c} \rightarrow p K^{-} $ &0.00 &0.03 \\ $ \Xi^{0}_{c}
\rightarrow n \bar K^{0} $ &-0.58 &0.04 \\ $ \Xi^{0}_{c} \rightarrow
\Lambda \pi^{0} $ &-0.79 &0.02 \\ $ \Xi^{0}_{c} \rightarrow \Sigma^{+}
\pi^{-} $ &0.00 &0.03 \\ $ \Xi^{0}_{c} \rightarrow \Sigma^{0} \pi^{0} $
&-0.89 &0.04 \\ $ \Xi^{0}_{c} \rightarrow \Sigma^{-} \pi^{+} $ &-0.99
&0.08 \\ $ \Xi^{0}_{c} \rightarrow \Xi^{-} K^{+} $ &-0.92 &0.06 \\ $
\Xi^{0}_{c} \rightarrow \Xi^{0} K^{0} $ &-0.40 &0.04 \\ $ \Xi^{0}_{c}
\rightarrow \Lambda \eta $ &$-0.89^1~(-0.88^2)$ &$0.02^1~(0.009^2)$ \\ $
\Xi^{0}_{c} \rightarrow \Lambda \eta' $ &$-0.99^1~(-0.99^2)$
&$0.04^1~(0.04^2)$ \\ $ \Xi^{0}_{c} \rightarrow \Sigma^{0} \eta $
&$-0.98^1~(-0.98^2)$ &$0.01^1~(0.02^2)$ \\ $ \Xi^{0}_{c} \rightarrow
\Sigma^{0} \eta' $ & $-0.99^1~(-0.99^2)$ &$0.006^1~(0.005^2)$ \\ & & \\
\hline \end{tabular} \\ $^1$For $\phi_{phy}=-10^0$,  $^2$For
$\phi_{phy}=-19^0$ \\

 \newpage \large  Table III(a) Branching ratios and asymmetries of CDS
  $(B_c \rightarrow BP)$ \\ decays  for $B(\Lambda^+_c \rightarrow
  \Xi^0K^+) = -16.52 \pm 2.19$  \baselineskip 24pt \begin{tabular}{ |
  c|c|c| }  \hline Decay & Asymmetry & $ Br\%$ $(\times
  tan^{4}_{\theta_{c}})$ \\ \hline & & \\ $ \Lambda^{+}_{c} \rightarrow
  p K^{0}$ &0.03 &3.15 \\ $ \Lambda^{+}_{c} \rightarrow n K^{+}$ &0.03
  &3.16 \\ & & \\ $ \Xi^{+}_{c} \rightarrow p\pi^{0} $ &-0.00 &1.41 \\ $
  \Xi^{+}_{c} \rightarrow n\pi^{+} $ &-0.00 &2.82 \\ $ \Xi^{+}_{c}
  \rightarrow \Lambda K^{+} $ &0.56 &0.54 \\ $ \Xi^{+}_{c} \rightarrow
  \Sigma^{+} K^{0} $ &-0.97 &4.39 \\ $ \Xi^{+}_{c} \rightarrow
  \Sigma^{0} K^{+} $ &-0.97 &2.19 \\ $ \Xi^{+}_{c} \rightarrow p \eta $
  &$0.52^1~(0.76^2)$ &$1.47^1~(1.15^2)$ \\ $ \Xi^{+}_{c} \rightarrow p
  \eta' $ &$-0.89^1~(-0.80^2)$ &$1.41^1~(1.68^2)$ \\  & & \\ $
  \Xi^{0}_{c} \rightarrow p \pi^{-} $ &-0.00 &0.79 \\ $ \Xi^{0}_{c}
  \rightarrow n \pi^{0} $ &-0.00 &0.40 \\ $ \Xi^{0}_{c} \rightarrow
  \Lambda K^{0} $ &0.56 &0.15 \\ $ \Xi^{0}_{c} \rightarrow \Sigma^{0}
  K^{0} $ &-0.97 &0.62 \\ $ \Xi^{0}_{c} \rightarrow \Sigma^{-} K^{+} $
  &-0.97 &1.24 \\ $ \Xi^{0}_{c} \rightarrow n \eta $ &$0.52^1~(0.76^2)$
  &$0.41^1~(0.32^2)$ \\ $ \Xi^{0}_{c} \rightarrow n \eta' $
  &$-0.89^1~(-0.80^2)$ &$0.39^1~(0.47^2)$ \\ & & \\ \hline \end{tabular}
  \\ $^1$For $\phi_{phy}=-10^0$,  $^2$For $\phi_{phy}=-19^0$ \\

\newpage Table III(b) Branching ratios and asymmetries of CDS $(B_c
\rightarrow BP)$ \\ decays  for $B(\Lambda^+_c \rightarrow \Xi^0K^+) =
+16.52 \pm 2.19$ \baselineskip 24pt \begin{tabular}{ | c|c|c| }  \hline
Decay & Asymmetry & $ Br\%$ $(\times tan^{4}_{ \theta_{c}})$ \\ \hline &
& \\ $ \Lambda^{+}_{c} \rightarrow p K^{0}$ &-0.19 &0.06 \\ $
\Lambda^{+}_{c} \rightarrow n K^{+}$ &-0.19 &0.06 \\ & & \\ $
\Xi^{+}_{c} \rightarrow p\pi^{0} $ &0.00 &1.41 \\ $ \Xi^{+}_{c}
\rightarrow n\pi^{+} $ &0.00 &2.82 \\ $ \Xi^{+}_{c} \rightarrow \Lambda
K^{+} $ &0.56 &0.54 \\ $ \Xi^{+}_{c} \rightarrow \Sigma^{+} K^{0} $
&-0.97 &4.39 \\ $ \Xi^{+}_{c} \rightarrow \Sigma^{0} K^{+} $ &-0.97
&2.19 \\ $ \Xi^{+}_{c} \rightarrow p \eta $ &$-0.89^1~(-0.72^2)$
&$1.88^1~(1.12^2)$ \\ $ \Xi^{+}_{c} \rightarrow p \eta' $
&$-0.94^1~(-0.96^2)$ &$3.05^1~(3.56^2)$ \\ & & \\  $ \Xi^{0}_{c}
\rightarrow p \pi^{-} $ &0.00 &0.79 \\ $ \Xi^{0}_{c} \rightarrow n
\pi^{0} $ &0.00 &0.40 \\ $ \Xi^{0}_{c} \rightarrow \Lambda K^{0} $ &0.56
&0.15 \\ $ \Xi^{0}_{c} \rightarrow \Sigma^{0} K^{0} $ &-0.97 &0.62 \\ $
\Xi^{0}_{c} \rightarrow \Sigma^{-} K^{+} $ &-0.97 &1.24 \\ $ \Xi^{0}_{c}
\rightarrow n \eta $ &$-0.89^1~(-0.72^2)$ &$0.53^1~(0.32^2)$ \\ $
\Xi^{0}_{c} \rightarrow n \eta' $ &$-0.94^1~(-0.96^2)$
&$0.86^1~(0.99^2)$ \\ & &  \\ \hline \end{tabular} \\ $^1$For
$\phi_{phy}=-10^0$,  $^2$For $\phi_{phy}=-19^0$ \\

 \newpage \large  Table IV Branching ratios of CF $(B_c \rightarrow DP)$
 decays.\\ \baselineskip 24pt \begin{tabular}{ | c|c| } \hline Decay & $
 Br\%$ \\ \hline &  \\ $ \Lambda^{+}_{c} \rightarrow \Delta^{++}K^{-}$
 &$0.70\pm 0.40^*$ \\ $ \Lambda^{+}_{c} \rightarrow \Delta^{+}\bar
 K^{0}$ &$0.23\pm 0.13$ \\ $\Lambda^{+}_{c} \rightarrow
 \Sigma^{*+}\pi^{0}$ &$0.46\pm 0.18$ \\ $\Lambda^{+}_{c} \rightarrow
 \Sigma^{*+}\eta$ &$0.21\pm 0.11^1~(0.14 \pm 0.10^2)$ \\
 $\Lambda^{+}_{c} \rightarrow \Sigma^{*0}\pi^+$ &$0.46\pm 0.18$ \\
 $\Lambda^{+}_{c} \rightarrow \Xi^{*0}K^+$ &$0.23\pm 0.09^*$ \\ & \\
 $\Xi^{+}_{c} \rightarrow \Sigma^{*+}\bar K^0$ &$0.00$ \\ $\Xi^{+}_{c}
 \rightarrow \Xi^{*0}\pi^+$ &$0.00$ \\ & \\ $\Xi^{0}_{c} \rightarrow
 \Sigma^{*+}K^-$ &$0.13\pm 0.07$ \\ $\Xi^{0}_{c} \rightarrow
 \Sigma^{*0}\bar K^0$ &$0.06\pm 0.04$ \\ $\Xi^{0}_{c} \rightarrow
 \Xi^{*0}\pi^0$ &$0.26\pm 0.10$ \\ $\Xi^{0}_{c} \rightarrow
 \Xi^{*0}\eta$ &$0.13\pm 0.06^1~(0.08 \pm 0.06^2)$ \\ $\Xi^{0}_{c}
 \rightarrow \Xi^{*-}\pi^+$ &$0.50\pm 0.20$ \\ $\Xi^{0}_{c} \rightarrow
 \Omega^- K^+$ &$0.45\pm 0.18$ \\  & \\ \hline \end{tabular} \\
 $^*$Input,  $^1$For $\phi_{phy}=-10^0$, $^2$For $\phi_{phy}=-19^0$ \\

  \newpage  \large Table V Branching ratios of CS  $(B_c \rightarrow
 DP)$ decays.\\ \baselineskip 24pt \begin{tabular}{ | c|c|c| }  \hline
 Decay & $ Br\%$ \\ \hline &  \\ $ \Lambda^{+}_{c} \rightarrow
 \Delta^{++}\pi^{-}$ &0.05 \\ $ \Lambda^{+}_{c} \rightarrow
 \Delta^{+}\pi^{0}$ &0.08 \\ $\Lambda^{+}_{c} \rightarrow
 \Delta^{+}\eta$ &$0.0005^1~(0.000001^2)$ \\ $\Lambda^{+}_{c}
 \rightarrow \Delta^{+}\eta'$ &$0.002^1~(0.002^2)$ \\ $\Lambda^{+}_{c}
 \rightarrow \Delta^{0}\pi^+$ &0.08 \\ $\Lambda^{+}_{c} \rightarrow
 \Sigma^{*+}K^0$ &0.006 \\ $\Lambda^{+}_{c} \rightarrow \Sigma^{*0}K^+$
 &0.01 \\ & \\ $\Xi^{+}_{c} \rightarrow \Delta^{++}K^{-}$ &0.12 \\
 $\Xi^{+}_{c} \rightarrow \Delta^{+}\bar K^{0}$ &0.04 \\ $\Xi^{+}_{c}
 \rightarrow \Sigma^{*+}\pi^{0}$ &0.07 \\ $\Xi^{+}_{c} \rightarrow
 \Sigma^{*+}\eta$ &$0.04^1~(0.03^2)$ \\ $\Xi^{+}_{c} \rightarrow
 \Sigma^{*+}\eta'$ &$0.007^1~(0.009^2)$ \\ $\Xi^{+}_{c} \rightarrow
 \Sigma^{*0}\pi^{+}$ &0.07 \\ $\Xi^{+}_{c} \rightarrow \Xi^{*0}K^{+}$
 &0.06 \\  & \\ $\Xi^{0}_{c} \rightarrow \Delta^{+}K^{-}$ &0.01 \\
 $\Xi^{0}_{c} \rightarrow \Delta^{0}\bar K^{0}$ &0.01 \\ $\Xi^{0}_{c}
 \rightarrow \Sigma^{*+}\pi^{-}$ &0.009 \\ $\Xi^{0}_{c} \rightarrow
 \Sigma^{*0}\pi^{0}$ &0.06 \\ $\Xi^{0}_{c} \rightarrow \Sigma^{*0}\eta$
 &$0.008^1~(0.004^2)$ \\ $\Xi^{0}_{c} \rightarrow \Sigma^{*0}\eta'$
 &$0.004^1~(0.004^2)$ \\ $\Xi^{0}_{c} \rightarrow \Sigma^{*-}\pi^+$
 &0.16 \\ $\Xi^{0}_{c} \rightarrow \Xi^{*0}K^0$ &0.004 \\ $\Xi^{0}_{c}
 \rightarrow \Xi^{*-}K^+$ &0.06 \\  & \\ \hline \end{tabular} \\ $^1$For
 $\phi_{phy}=-10^0$,  $^2$For $\phi_{phy}=-19^0$ \\

 \newpage  Table VI   Branching ratios  of CDS $(B_c \rightarrow DP)$
decays. \\ \baselineskip 24pt  \begin{tabular}{ | c|c|c| }  \hline Decay
& $ Br\%$ $(\times tan^{4}_{\theta_{c}})$ \\ \hline & \\ $
\Lambda^{+}_{c} \rightarrow \Delta^{+} K^{0}$ &0 \\ $ \Lambda^{+}_{c}
\rightarrow \Delta^{0} K^{+}$ &0 \\ & \\ $ \Xi^{+}_{c} \rightarrow
\Delta^{++} \pi^{-}$ &3.00 \\ $ \Xi^{+}_{c} \rightarrow \Delta^{+}
\pi^{0}$ &4.79 \\ $ \Xi^{+}_{c} \rightarrow \Delta^{0} \pi^{+}$ &4.39 \\
$ \Xi^{+}_{c} \rightarrow \Sigma^{*0} K^{+}$ &0.99 \\ $ \Xi^{+}_{c}
\rightarrow \Sigma^{*+} K^{0}$ &0.45 \\ $ \Xi^{+}_{c} \rightarrow
\Delta^{+} \eta$ &$0.03^1~(0.0001^2)$ \\ $ \Xi^{+}_{c} \rightarrow
\Delta^{+} \eta'$ &$0.38^1~(0.39^2)$ \\ & \\ $ \Xi^{0}_{c} \rightarrow
\Delta^{+} \pi^-$ &0.28 \\ $ \Xi^{0}_{c} \rightarrow \Delta^{-} \pi^+$
&3.73 \\ $ \Xi^{0}_{c} \rightarrow \Delta^{0} \pi^0$ &1.36 \\ $
\Xi^{0}_{c} \rightarrow \Sigma^{*-} K^+$ &0.56 \\ $ \Xi^{0}_{c}
\rightarrow \Sigma^{*0} K^0$ &0.06 \\ $ \Xi^{0}_{c} \rightarrow
\Delta^{0} \eta$ &$0.01^1~(0.00002^2)$ \\ $ \Xi^{0}_{c} \rightarrow
\Delta^{0} \eta'$ &$0.11^1~(0.11^2)$ \\ &  \\ \hline \end{tabular} \\
\hskip 0.5cm $^1$For $\phi_{phy}=-10^0$,  $^2$For  $\phi_{phy}=-19^0$ \\
\end{center}

 \newpage \large 
 \end{document}